\def\addrParma{Dipartimento di Fisica e Scienze della Terra, Universit\`a degli Studi di Parma, Viale G.\ Usberti 7A, I-43100 Parma, Italy}
\def\LSCO{La$_{5/3}$Sr$_{1/3}$CoO$_4$}

\def\Cot{Co$^{3+}$}
\def\Cod{Co$^{2+}$}

\documentclass[prb,twocolumn,showpacs,amsmath,amssymb,superscriptaddress]{revtex4}
\usepackage{graphicx}
\usepackage{dcolumn}
\usepackage{bm}
\usepackage{epsfig}
\begin{document}

\title{Stripe disorder and dynamics in the hole-doped
  antiferromagnetic insulator La$_{5/3}$Sr$_{1/3}$CoO$_{4}$}

\author{T. Lancaster}
\affiliation{Centre for Materials Physics, Durham University, South Road, 
Durham, DH1 3LE, United Kingdom}
\author{S.R Giblin}
\affiliation{School of Physics and Astronomy, Cardiff
  University, Queen's Buildings, The Parade, Cardiff, CF24 3AA, United
  Kingdom}
\author{G. Allodi}
\affiliation{\addrParma}
\author{S. Bordignon}
\affiliation{\addrParma}
\author{M. Mazzani}
\affiliation{\addrParma}
\author{R. De Renzi}
\affiliation{\addrParma}
\author{P.G. Freeman}
\altaffiliation{Current address: Laboratory for Quantum Magnetism, Ecole Polytechnique F\'{e}d\'{e}rale de Lausanne, CH-1015 Lausanne, Switzerland}
\affiliation{Helmholtz-Zentrum Berlin f\"{u}r Materialien und Energie, Hahn-Meitner-Platz 1, DE-14109 Berlin, Germany}
\affiliation{Institut Laue-Langevin, BP 156, 38042 Grenoble Cedex 9, France}
\author{P.J. Baker}
\author{F.L. Pratt}
\affiliation{ISIS Facility, Rutherford Appleton Laboratory, Chilton, 
Oxfordshire OX11 0QX, UK}
\author{P. Babkevich}
\altaffiliation{Current address: Laboratory for Quantum Magnetism, Ecole Polytechnique F\'{e}d\'{e}rale de Lausanne, CH-1015 Lausanne, Switzerland}
\affiliation{Oxford University Department of Physics, Clarendon
  Laboratory, Parks Road, Oxford, OX1 3PU, United Kingdom}
\author{S.J. Blundell}
\author{A.T. Boothroyd}
\author{J.S. M\"oller}
\author{D. Prabhakaran}
\affiliation{Oxford University Department of Physics, Clarendon Laboratory, 
Parks Road, Oxford, OX1 3PU, United Kingdom}

\date{\today}

\begin{abstract}
We investigate the magnetic ordering and
dynamics of the stripe phase of La$_{5/3}$Sr$_{1/3}$CoO$_{4}$, a
material shown to have an hour-glass magnetic excitation spectrum. A
combination of muon-spin relaxation, nuclear magnetic resonance and magnetic
susceptibility
measurements strongly suggest that the physics is determined by a
partially disordered configuration of charge
and spin stripes whose frustrated magnetic degrees of freedom are
 dynamic at high temperature and which undergo an ordering
 transition around 35~K with coexisting dynamics that
freeze out in a glassy manner as the temperature is further reduced.
\end{abstract}
\maketitle

\begin{figure}
\begin{center}
\epsfig{file=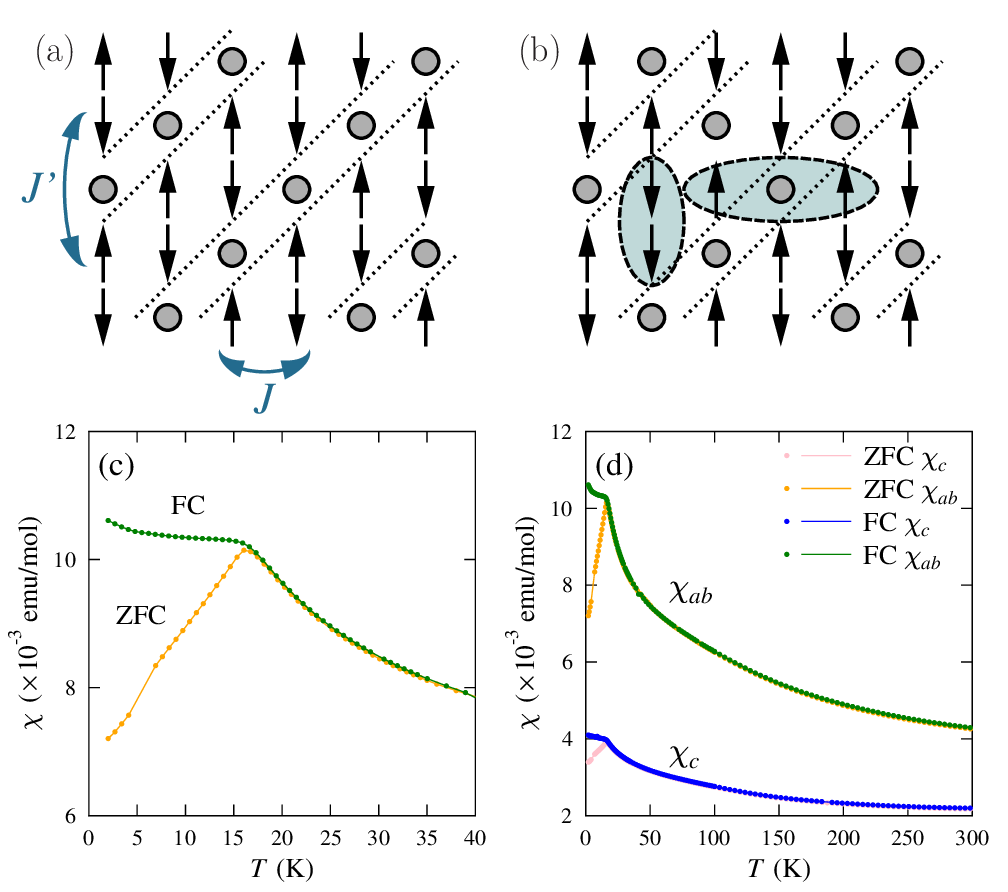,width=7cm}
\caption{(a) Perfect stripe order in the CoO$_{2}$ planes. (Circles
represent non-magnetic Co$^{3+}$, arrows represent
$S=3/2$ Co$^{2+}$ spins). 
(b) Disordering the charges leads to unfavorable exchange interactions
between spins (shaded). 
(c) and (d) Magnetic susceptibility as a function of temperature for
field cooled and zero-field cooled protocols for two crystal orientations.
\label{stripes}}
\end{center}
\end{figure}

The hour-glass spectrum of spin excitations observed using
inelastic neutron scattering (INS) in the
cuprate superconductors
\cite{arai,bourges,hayden,reznik,stock,hinkov,fauque,xu,christensen,vignolle,matsuda,lipscombe,tranquada2} 
has been linked to the occurrence
of alternating patterns of spin and charge stripes in the
copper oxide planes.\cite{tranquada}
Although many cuprates exhibit hour-glass dispersion and
show no evidence for stripe order, the discovery of such an
excitation spectrum in stripe-ordered cobaltate 
 materials provided strong evidence that the hour-glass dispersion results
from short-range stripe correlations.\cite{boothroyd,gaw} 
The main features of the hour-glass spectrum 
can be reproduced by the spin-wave spectrum of perfectly ordered,
weakly coupled antiferromagnetic (AFM) stripes [Fig.~\ref{stripes}(a)] with phenomenological
broadening.\cite{boothroyd} Recently, however, a more detailed
agreement has been obtained by a spin-wave calculation based on a
stripe model that explicitly incorporates quenched disorder in the
charge degrees of freedom and whose magnetic moments, through
frustration, may be described in terms of cluster-glass behaviour at
low temperature.\cite{andrade} (We also note here the more recent observation
of an hour-glass spectrum in another cobaltate material, within the
checkerboard charge-ordered regime, where stripes have not been
observed and where an alternative origin for the spectrum is suggested.\cite{drees})

The wider implications of the link between stripes and hour-glass
dispersion, on the cuprates in particular, has motivated this
investigation
into a hitherto unexplored aspect of the stripe phase dynamics in La$_{5/3}$Sr$_{1/3}$CoO$_{4}$.
The study of very low-frequency excitations related to stripes, such as their slow collective
motion, is outside the scope of inelastic neutron scattering.
We have therefore selected
muon-spin
relaxation ($\mu^{+}$SR) and nuclear magnetic resonance
(NMR)
as probes
of this behavior.
We note that, in contrast to muon and NMR spectroscopy, the previous
INS measurements were insensitive to 
fluctuations on timescales much slower than $\hbar/\Delta E \approx10^{-11}$s (where 
$\Delta E \approx 1$~meV is the energy scale of the resolution of the
measurement)
and so fluctuations on timescales longer than this appeared static.
INS therefore took a ``snap-shot'' of the behaviour
compared to $\mu^{+}$SR and NMR measurements whose 
characteristic time scale is set by the respective gyromagnetic ratios
of the muon ($\gamma_{\mu} = 2 \pi\times 135.5$~MHz
T$^{-1}$)  and  the nuclei being interrogated. 
In this paper we show that $\mu^{+}$SR and NMR find dynamics, magnetic
ordering (around 35~K) and a freezing 
of dynamically fluctuating moments (around 20~K) which is consistent with a picture of
partially disordered stripes whose dynamics are frozen out as the temperature is
lowered. 

The La$_{2-x}$Sr$_{x}$CoO$_{4}$ system is based around well isolated, square layers of
CoO$_{2}$ and is
isostructural to the `214' family of cuprates, which includes
La$_{2-x}$Sr$_{x}$CuO$_{4}$.
In the parent ($x=0$) compound, commensurate AFM order has been reported
below\cite{yamada} $T=275$~K. Hole doping of the material involves exchanging
Sr for La, resulting in the donation of positive charges to the CoO layers.
For a doping of $x>0.3$, magnetic order is modulated at 45$^{\circ}$ to the CoO bonds 
which is attributable to the
self-organization of holes into arrays of charged stripes, 
which creates antiphase domain walls in the antiferromagnetic order.
\cite{cwik}
The stripe order in La$_{5/3}$Sr$_{1/3}$CoO$_{4}$
involves diagonal lines of non-magnetic $S=0$ Co$^{3+}$ ions separated by bands of antiferromagnetically
aligned Co$^{2+}$ $S=3/2$ ions, with intra- ($J$) and
inter-stripe ($J'$) antiferromagnetic exchange couplings  as shown in
Fig.~\ref{stripes}(a). It is currently assumed that the charge
ordering of the Co ions sets in at a temperature $T_{\mathrm{CO}}$
well above room temperature, while magnetic Bragg peaks
are observed in neutron diffraction below\cite{boothroyd2} $\sim 100$~K.
The correlation lengths of the magnetic order
were estimated to be $\xi = 10$~\AA\ parallel to the stripes, and $\xi
= 6.5$~\AA\ perpendicular to them, pointing to a short-ranged
character to the magnetism, which is unlikely to show conventional
critical dynamics.
\cite{boothroyd}
Disordered stripes may be formed by rearranging the charges of the configuration shown in Fig~\ref{stripes}(a). 
Imperfections in the charge order are expected to be static at
temperatures which are low compared to $T_{\mathrm{CO}}$ owing to the
 insulating nature of the material, while dynamic fluctuations should
be expected in the spin degrees of freedom below room temperature. 
As shown in Fig~\ref{stripes}(b), the introduction of
charge disorder 
frustrates the 
antiferromagnetic coupling between
spins, and this is the origin of the proposed cluster-glass behavior in this system. \cite{andrade}


\begin{figure}
\begin{center}
\epsfig{file=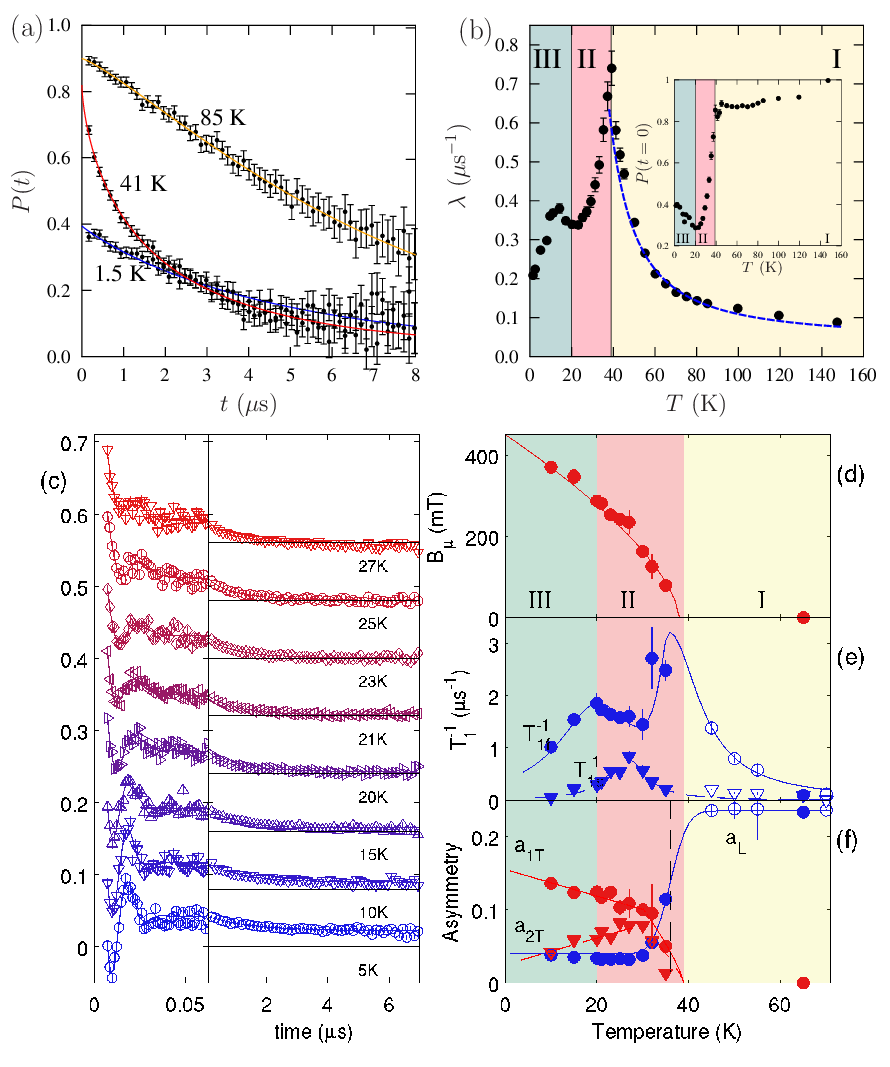,width=7cm}
\caption{
 (a) Example zero field $\mu^{+}$SR spectra measured at ISIS. 
(b) ISIS relaxation rate $\lambda$ with a fit to an activated behaviour at high
temperatures (see main text); {\it inset}: Initial muon polarization $P(0)$.
(c) Data measured at S$\mu$S showing oscillations at early times.
(d) The larger of the two  precession frequencies as a function of temperature. (e)
Evolution of longitudinal relaxation rates $1/T_{1}$. (f) Amplitudes
of the transverse and longitudinal components. 
\label{muondata}}
\end{center}
\end{figure}

To investigate spin ordering and dynamics
$\mu^{+}$SR measurements \cite{steve,yaouanc} were made
with the initial
muon spin directed along the $c$-axis of the material, so that the
muon spins are initially perpendicular to the CoO$_{2}$ layers. 
To survey the slow dynamics of the system
 $\mu^{+}$SR data were measured using the EMU instrument at the ISIS
 facility [Fig.~\ref{muondata}(a)], where time resolution limitations
do not allow us to resolve oscillations or relaxation rates $\gtrsim 10$~MHz. Instead, we
see a decrease in the relaxing amplitude $A_{\mathrm{rel}}$ as the
material is cooled below 40~K [inset, Fig.~\ref{muondata}(b)]. This is because the local
field in the ordered state strongly depolarizes the muon
spin if local field components are perpendicular to the
initial muon-spin polarization 
removing relaxing asymmetry from the
spectrum.
In view of the limitations caused by the ISIS time resolution, 
we parametrize the muon asymmetry with a stretched exponential function
$
a(t) = a(0) P(t) = A_{\mathrm{rel}} {\rm e}^{-(\lambda t)^{\beta}}+a_{\mathrm{bg}}
\label{fitfunc}
$
where $P(t)$ is the muon ensemble spin polarization, $a_{\mathrm{bg}}$
is a background contribution.
We expect that\cite{yaouanc}, in the fast fluctuation limit, the muon relaxation rate 
is determined by the relation $\lambda \propto \Delta^{2}\tau$,
where $\Delta^{2} = \gamma^{2}_{\mu}\langle B_{i}^{2} \rangle$
 and $\tau$ is the correlation time. 
We note that no features are observed around 100~K, where 
magnetic Bragg peaks become resolvable. 
Three regions in temperature are apparent in the behavior. On cooling
in the paramagnetic region I ($T>40$~K)
the relaxation rate $\lambda$ increases slowly, then more
rapidly below $\approx 70$~K before peaking at $T=39$~K [Fig.~\ref{muondata}(b)]. 
The initial polarization $P(t=0)$ falls slowly across region
I. On entering region II ($20 \lesssim T \leq 39$~K)
$P(0)$ drops rapidly and $\lambda$ decreases, reaching a
minimum at 20~K. Finally, in region III 
($T \lesssim 20$~K) $P(0)$ increases and there is another peak
in $\lambda$  centred around 15~K. 

A more complete picture of the muon response is obtained from measurements 
made using the DOLLY instrument at the Swiss Muon Source (S$\mu$S) where
shorter time scales may be resolved [Fig.~\ref{muondata}(c-f)]. 
In region I we observe rapidly and slowly relaxing contributions with
relaxation rates rates $T_{1\mathrm{f}}^{-1}$ and $T_{1\mathrm{s}}^{-1}$ respectively.
In regions II and III
 ($T \lesssim 39$~K)
we observe oscillations in the muon spectra at early times ($t\leq 0.05~\mu$s),
characteristic of a static local magnetic field at the 
muon stopping site, causing a coherent precession of the
spins of those muons with a component of their spin polarization
perpendicular to this local field. The frequency of the oscillations is given by
$\nu_{i} = \gamma_{\mu} B_{i}/2 \pi$, where $B_{i}$
is the average magnitude of the local magnetic field at the $i$th muon
site.
In this temperature regime, two relaxing terms account for the
longitudinal component of the asymmetry 
and two oscillatory terms
provide the transverse asymmetry component.\cite{suppl}
Data in Fig.~\ref{muondata}(c) and (f) show
the presence of these transverse terms in the early-time oscillations 
and longitudinal terms in the late-time relaxation. 
The ratio of the two local fields was found to be constant\cite{suppl} and
Fig~\ref{muondata}(d) shows the largest of them as a function of temperature,
which is found to vanish at $T_{\mathrm{N}} \approx 35$~K. 
Fig.~\ref{muondata}(f) shows that the total amplitude of the
transverse terms, 
$a_{1\mathrm{T}}+a_{2\mathrm{T}}$, vanishes above $T_{\mathrm{N}}$, where the total
longitudinal amplitude $a_{\mathrm{L}} =
a_{\mathrm{s}}+a_{\mathrm{f}}$ recovers the 
full 
asymmetry
(the distinction between longitudinal and transverse being lost above
$T_{\mathrm{N}}$, where the spin system recovers rotational
invariance). On the strength of $\mu^{+}$SR, we are therefore able
to identify 
a transition to magnetic order around
$T_{\mathrm{N}}=35$~K, 
broadened by significant inherent static disorder, leading to a Gaussian width
$\Delta T_{\mathrm{N}}$ approx 5 K.

The presence of
significant static disorder is confirmed by 
the broad distribution of local fields providing the fast damping of
the oscillations [Fig.~\ref{muondata}(c)], 
The dynamics driving the two longitudinal relaxation rates also deviate
from the behaviour expected 
for a homogeneous magnetically ordered phase,
for which we expect monotonically increasing relaxation rates up to a critical divergence at $T_{\mathrm{N}}$.\cite{DeRenzi}
In contrast, Fig.~\ref{muondata}(e) suggests not only a peak in
$T^{-1}_{1\mathrm{f}}$ around $T_{\mathrm{N}}$  significantly broadened in the presence of strong disorder (see e.g.\ Ref.~\onlinecite{ybco}), 
but also shows
additional peaks in $T^{-1}_{1\mathrm{f}}$ and $T^{-1}_{1\mathrm{s}}$
around the crossover between region II and III near 20~K.
As argued below, these results are consistent with 
region II, in addition to showing magnetic order, also displaying
significant dynamics which freeze out on cooling, with correlation
times starting to become longer than the $\mu^{+}$SR 
time window in the more static and ordered region III. This implies that the border between region II and III is actually a blurred crossover.


To probe the slow dynamics of both charge and magnetic degrees of freedom,
NMR measurements were made on a single crystal sample of La$_{5/3}$Sr$_{1/3}$CoO$_{4}$
with external field 
applied in the $ab$ plane.
We expect that only $^{59}$Co nuclei in spinless \Cot\ ions will be
detected, since for a magnetic Co$^{2+}$ ion, the $^{59}$Co nucleus experiences an instantaneous  
hyperfine field  of order 10~T$\mu_{\mathrm{B}}^{-1}$, 
whose fluctuations 
lead to very fast nuclear relaxation.\cite{freeman_watson} 

\begin{figure}[b]
\begin{center}
\includegraphics[width=7cm]{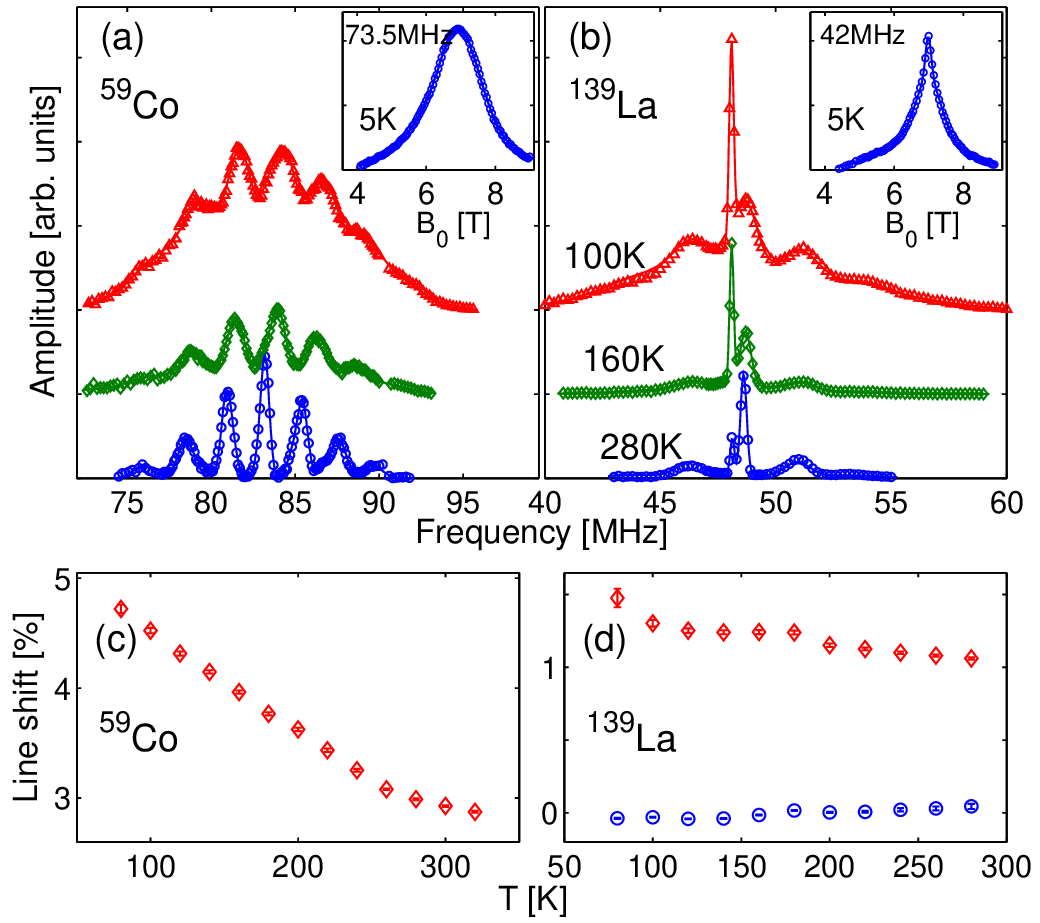}
\caption{(Color online) \label{fig:spectral}
Top: frequency-swept (a) $^{59}$Co and  (b) $^{139}$La  NMR spectra in 
$B_{\mathrm{app}}=8$~T at three selected temperatures. Insets: field-swept spectra at 
5~K. Bottom: shifts 
of 
(c) the central line of $^{59}$Co,
and (d) the central doublet of $^{139}$La as a function of temperature. 
}
\end{center}
\end{figure}

Example frequency-swept spectra from probe nuclei $^{59}$Co and $^{139}$La, 
measured in $B_{\mathrm{app}}=8$~T, are shown in
Fig.~\ref{fig:spectral}(a-b) (field-swept data shown inset).
At room temperature, the $^{59}$Co spectrum consists of a
well-resolved line septet, 
originating from 
the nuclear $I=7/2$ spin, 
split by the quadrupole coupling with the local electric field gradient (EFG).  
The central line, unperturbed to first order by the EFG, exhibits a
sizeable shift 
with respect to the  $^{59}$Co reference
[Fig.~\ref{fig:spectral}(c)].
The hyperfine coupling 
of $^{59}$Co is estimated as
$\approx 5$~T  from the comparison of 
the shift with the magnetic susceptibility (see below)
 a value compatible with a transferred hyperfine contribution from neighbouring Co$^{2+}$ onto the non magnetic Co$^{3+}$,
confirming the
localization of holes in the layers. 
The $^{139}$La ($I=7/2$) signal shows similar
quadrupolar patterns, with higher-order satellites smeared out by EFG 
inhomogeneities which are larger at the La site. The  central line is split
into a doublet by magnetic interactions 
  resulting
from the occupancy of the nearest neighbor Co site by high-spin \Cod\ or spinless \Cot.
The majority $^{139}$La  peak exhibits larger and temperature-dependent shifts, while the smaller 
shift of the minority peak is nearly temperature-independent  [Fig.~\ref{fig:spectral}(d)]. 

On cooling, the spectra broaden, 
most dramatically seen in the $^{59}$Co signal which shows
a broad shoulder 
superimposed on the quadrupole septet at 100~K [Fig.~\ref{fig:spectral}(a)].
Such 
broadening reflects the onset of significant magnetic correlations below approximately the 
same temperature as the onset of magnetic Bragg peaks in neutron scattering.  \cite{boothroyd2}
It is likely that the static order here is  short-ranged 
and results from the large applied field $B_{\mathrm{app}}$.
Line broadening continues on cooling without abrupt
changes
with almost
featureless spectra observed for  $T  <10 $~K [Fig.~\ref{fig:spectral} (insets)].

\begin{figure}[t] 
\begin{center}
\includegraphics[width=7cm]{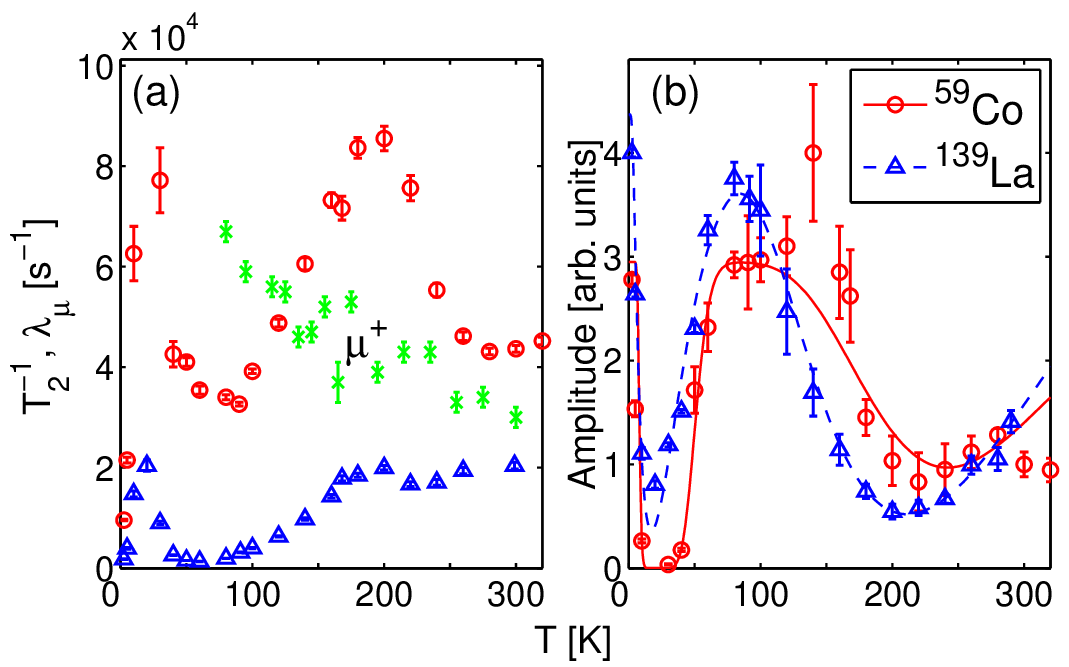}
\caption{(Color online) \label{fig:NMRrelax}
(a) Evolution of spin-spin relaxation rates  $T_2^{-1}$ with $T$  for $^{59}$Co (circles),
$^{139}$La  (triangles) and muon relaxation rates $\lambda$ (crosses). (b) Integrated spectral amplitudes, corrected
for $T_2^{-1}$
relaxation as a function of temperature.}
\end{center}
\end{figure}

The key result from our NMR is that the signal amplitude is partially lost
(or wiped out) in two temperature intervals: above 100~K and, more severely, in the 10--40 K
range. We note that
the measured spin-spin relaxation rates $T_2^{-1}$ 
 [Fig.~\ref{fig:NMRrelax}(a)]
exhibit two maxima: a broad peak round 200~K, and a 
sharper one at $T~\approx 20$~K.
Where the relaxation rates of the two nuclei develop peaks [Fig.~\ref{fig:NMRrelax}(a)]
the corresponding amplitudes, corrected for the initial deadtime
and the Curie temperature dependence are severely
reduced [Fig.~\ref{fig:NMRrelax}(b)], implying that they correspond to residual signals. 
Faster spin-spin relaxations manifest themselves as sizeable missing
fractions (i.e.\ a signal wipeout), coinciding with the $T_2^{-1}$ peaks.
A wipeout reflects a  broad distribution of relaxation rates,
where signal components with $T_2$ much shorter than the 
instrumental dead time
($\approx 10~\mu$s) are lost.
A model of the NMR wipeouts which assumes a log-normal
distribution of activation energies\cite{suppl} suggests a characteristics 
 energy scales for the two distict observed $T_{2}^{-1}$ peaks at 
$E_{\mathrm{a}} \approx 1100$~K and 70~K.


Both peaks are indicative of very slow dynamic excitations of the
stripes. 
These slow excitations may involve  either separate charge and spin
degrees of
 freedom or their interplay. We are able to discriminate this aspect
 by 
comparing the relaxation of the two nuclei with that of the muons. 
Both nuclei are sensitive to slow dynamics of spin and charge as well, since they are coupled to the  EFG via their quadrupole moment. 
The fast relaxations of $^{59}$Co and $^{139}$La above 100~K have 
no counterpart in our $\mu^{+}$SR data [Figs.~\ref{fig:NMRrelax}(a)], 
and therefore must be due to 
EFG fluctuations, to which $I=1/2$ muons are not sensitive.
The excitations underlying such EFG fluctuations
most likely consist of thermally assisted hopping of holes across 
stripes, apparently taking place well below the charge ordering temperature. 
Such charge motion slows down on cooling, down to a complete 
freezing at temperatures of the order of 100~K, at which the full NMR 
signal amplitude is recovered. 
The origin of the low temperature wipeout is 
magnetic, as described below.

Measurements of DC magnetic
susceptibility $\chi$, made using a Quantum design MPMS [Fig.~\ref{stripes}(c) and (d)]
show no strong features around $100$~K [Fig.~\ref{stripes}(d)]
where magnetic Bragg peaks start to appear in neutron diffraction. 
Susceptibilities measured after zero-field cooling
and field cooling protocols (measuring field
of $B=100$~mT) show a gradual increase upon cooling before splitting
below 
$T=18$~K [Fig~\ref{stripes}(c)]. This splitting is typical of a freezing of the
magnetic moments
at low temperature and reflects the fact that, after cooling in ZF
to a region where the spins are statically frozen, we obtain a configuration which is less susceptible to magnetization 
than occurs if the frozen state is achieved in an applied field. 
AC susceptibility measurements were attempted on this system  but a
response could not be resolved.

Taken together, the $\mu^{+}$SR, NMR and  magnetic susceptibility
results allow an insight into the behavior of the
stripe ordering and dynamics in La$_{5/3}$Sr$_{1/3}$CoO$_{4}$.
{\it Region I}: The snap-shot taken by neutron diffraction indicates that
ordered stripes have formed and are static over intervals $\tau > 10^{-11}$s, at
temperatures below 100~K. 
However, the $\mu^{+}$SR suggests that at these high temperatures spins within the stripes are rapidly
fluctuating on the muon ($\mu$s)  time scale. 
A crude estimate suggests that the observed muon relaxation rate, of
order $0.15~\mu$s$^{-1}$, is consistent with a local magnetic field of around
150~mT fluctuating with correlation times of order 10$^{-11}$~s.
A decrease in temperature causes these fluctuations to slow down, increasing
$\tau$ and hence $\lambda$. Assuming an activated behaviour of the
fluctuations in this region  [Fig.~\ref{muondata}(b)] with $\tau = \tau_{\infty} e^{E_{\mathrm{a}}/ T}$, 
we obtain an activation energy $E_{\mathrm{a}}\approx 100$~K, in
reasonable agreement with the energy scale identified from NMR. 
{\it Region II}: Temperature $T \approx 40$~K marks the 
point where stripe spins 
lock together on the muon timescale and we enter a regime of magnetic order. 
As the system is cooled through the  $40$--$20$~K region, 
slow dynamics remain significant, leading to the low-temperature NMR
wipeout as well as   to the muon $T_{1\mathrm{f}}^{-1}$ peak.
Assuming  a simple Lorentzian spectral density, the
maximum $T^{-1}$ relaxation rate seen in the $\mu^{+}$SR data  around
20~K [Fig.\ref{muondata}(e)]
occurs \cite{suppl} when $\tau^{-1}(T)=\gamma_{\mu} B$, where $B$ is the
internal magnetic field. Taking $B=300$~mT when $T_{1}^{-1}=2 \times 10^6$~s$^{-1}$ 
 implies $\tau(T\approx 20~\mathrm{K}) \approx 4$~ns and $\Delta \approx 30$~$\mu$s$^{-1}$. 
For comparison, the lognormal distribution of correlation times obtained
from the $^{59}$Co wipe-out in NMR [Fig.~4 \cite{suppl}]  predicts a range of
correlation times 
$5 \times 10^{-10} < \tau < 8 \times 10^{-9}$~s for $T=20$~K, in good agreement.
{\it Region III}: Peaks in the muon $\lambda$ and $T^{-1}$ and the increase in
$A_{\mathrm{rel}}$ below 20~K, suggest a crossover to a more static configuration 
pointing to the dynamics of the stripes freezing out below 20~K. 
This low temperature freezing accounts for  the splitting of the ZFC and FC $\chi$ data and the reappearance of the NMR signal.


We may compare our results to others measured on
stripe-ordered nickelate and cuprate systems.\cite{klauss} Muon measurements on both La$_{5/3}$Sr$_{1/3}$NiO$_{4}$
and the La$_{1.8-x}$Eu$_{0.2}$Sr$_{x}$CuO$_{4}$
system show well-resolved oscillations setting in below ordering temperatures
$T_{\mathrm{N}}\approx 200$~K, with signatures of stripe dynamics seen
in anomalies in the precession frequencies and/or the longitudinal relaxation rate.\cite{klauss} In
La$_{5/3}$Sr$_{1/3}$NiO$_{4}$, where NMR measurements are  suggestive of glassy
stripe dynamics,\cite{yoshinari} the muon and NMR $T_{1}^{-1}$ both decrease monotonically with decreasing
temperature for $T <200$~K in contrast to what we report
here. However, the correlation length in the nickelate system,
estimated \cite{yoshizawa} as $\xi \gtrsim 100$~\AA,
is far larger than that of \LSCO\  making it hard to see how similar quenched charge
disorder drives the physics in that case. For the Eu-containing cuprate system, where NMR shows a
low temperature wipeout and there is evidence for a distribution of
timescales,\cite{curro} it is possible that the broad peak\cite{klauss} around 7~K measured in
the $\mu^{+}$SR $1/T_{1}$ has a similar origin to the behaviour we observe.

In conclusion, the stripes that dominate the low-energy physics of
\LSCO\ order on the muon time-scale around 35~K, but display
slow dynamics that freeze
out of spin dynamics upon cooling below 20~K. 
The distribution of activated spin correlation times detected by both
muons and NMR 
indicate glassy behaviour which is most probably due to the inherent
frustration introduced 
by charge disorder in the magnetic couplings.
 However the presence of a high temperature wipeout in the NMR,
which has no analogue in $\mu^{+}$SR, indicates that this charge quenching is
not complete. The influence of such stripe phase dynamics on
the physics of the cuprates remains an open and intriguing question \cite{andrade}.

Part of this work was carried out at the S$\mu$S, Paul Scherrer
Institut, Switzerland and at the STFC ISIS facility, UK and we are
grateful for the provision of beamtime.
We thank R. Scheuermann and A. Amato for technical support. 
We thank EPSRC (UK) for financial support.

\end{document}



\title{Supplemental Information}
\author{T. Lancaster}
\affiliation{Centre for Materials Physics, Durham University, South Road, 
Durham, DH1 3LE, United Kingdom}
\author{S.R Giblin}
\affiliation{School of Physics and Astronomy, Cardiff
  University, Queen's Buildings, The Parade, Cardiff, CF24 3AA, United
  Kingdom}
\author{G. Allodi}
\affiliation{\addrParma}
\author{S. Bordignon}
\affiliation{\addrParma}
\author{M. Mazzani}
\affiliation{\addrParma}
\author{R. De Renzi}
\affiliation{\addrParma}
\author{P.G. Freeman}
\altaffiliation{Current address: Laboratory for Quantum Magnetism, Ecole Polytechnique F\'{e}d\'{e}rale de Lausanne, CH-1015 Lausanne, Switzerland}
\affiliation{Helmholtz-Zentrum Berlin f\"{u}r Materialien und Energie, Hahn-Meitner-Platz 1, DE-14109 Berlin, Germany}
\affiliation{Institut Laue-Langevin, BP 156, 38042 Grenoble Cedex 9, France}
\author{P.J. Baker}
\author{F.L. Pratt}
\affiliation{ISIS Facility, Rutherford Appleton Laboratory, Chilton, 
Oxfordshire OX11 0QX, UK}
\author{P. Babkevich}
\altaffiliation{Current address: Laboratory for Quantum Magnetism, Ecole Polytechnique F\'{e}d\'{e}rale de Lausanne, CH-1015 Lausanne, Switzerland}
\affiliation{Oxford University Department of Physics, Clarendon
  Laboratory, Parks Road, Oxford, OX1 3PU, United Kingdom}
\author{S.J. Blundell}
\author{A.T. Boothroyd}
\author{J.S. M\"oller}
\author{D. Prabhakaran}
\affiliation{Oxford University Department of Physics, Clarendon Laboratory, 
Parks Road, Oxford, OX1 3PU, United Kingdom}

\date{\today}







\maketitle
\section{A Model of the NMR wipeout}
\label{sec:intro}
A simple model of the NMR wipe-out which is used to identify the
relevant energy scales in the main paper
is described below. 

\subsection{ Assumptions of the model} 
The model assumes the following:

\begin{itemize}
\item 
The relaxation rate $T_{2}^{-1}$ is given by
\begin{equation} %
T_2^{-1}=\Delta\omega^2\,\tau, 
\end{equation} %
where $\Delta \omega^{2}$ is the second moment of the frequency
distribution and $\tau$ the correlation time. 

Strictly, this is correct in the motional narrowing (or fast fluctuation) limit.
From the experimentally observed values of $T_1$ and $T_2$,  we
find that this regime is apparently achieved on the high-temperature
side of the $A(T)$ curve, down to well inside the wipeout region. 
Within the model, the exact $T_2(\tau)$ dependence, on the other hand, is irrelevant for the low-temperature signal recovery (see below).
\item An activated correlation time
\begin{equation}
\label{eq:activated}
\tau = \tau_{\infty}\exp(E/T),
 \end{equation}
with the activation energy $E$  distributed according to a log-normal 
distribution:
\begin{equation}
\label{eq:distro}
p(E)= \frac{1}{E\sqrt{2\pi}\sigma} \exp \left ({-\frac{\log \left (\frac{E}{E_a} \right )^2 }{2\sigma^2} }\right).
\end{equation}

\item 
A minimum observable relaxation time $T_{2\,\mathrm{min}}$, 
so that signal components with $T_2 < T_{2\,\mathrm{min}}$ are completely lost, 
while  $T_2 > T_{2\,\mathrm{min}}$ fully contribute to the signal
amplitude.
We estimate
$T_{2\,\mathrm{min}}$ 
of the order of 5-7~$\mu$s.

\item The finite time window $t_{\mathrm{w}}$ of the spin echo experiment, corresponding to the excitation-detection sequence ($P_1$-delay-$P_2$-delay-acquisition), of 
the order of 20-30~$\mu$s.

\item In the nearly-static limit $\tau \approx t_{\mathrm{w}}$, a
  strong collision model applies.
Fluctuation events occur with a probability $1/\tau$ per unit time,
if they occur they completely destroy the spin-echo.
The probability for no collision event over $t_{\mathrm{w}}$ is then $\exp(-t_{\mathrm{w}}/\tau)$.
  
\end{itemize}

\subsection{NMR amplitude vs.\ $\tau$}
An NMR signal is observed if  $T_2(\tau) >
T_{2\,\mathrm{min}}$ 
or if no fluctuation takes
place over 
a time $t_{\mathrm{w}}$. Hence, we write the amplitude as
\begin{equation}
\label{eq:amp}
A(\tau) = \left (1-e^{-\frac{t_{\mathrm{w}}}{\tau}} \right
)\Theta\left ( T_2(\tau) - T_{2\,{\mathrm{min}}} \right ) + e^{-\frac{t_{\mathrm{w}}}{\tau}},
\end{equation}
where $\Theta$ is the step function: $\Theta(x)= 1$ for $x\ge 0$,  $\Theta(x)= 0$ for $x< 0$.
We also define 
\begin{eqnarray} 
\label{eq:defs}
\tau_{\mathrm{R}} & \equiv &\frac{1}{\Delta\omega^2\,T_{2\,{\mathrm{min}}}}, \nonumber \\
\alpha & \equiv &\ln\left( \frac{T_{2\,\infty}}{T_{2\,{\mathrm{min}}}}\right ) =
\ln\left( \frac{1}{\Delta\omega^2\tau_\infty T_{2\,{\mathrm{min}}}}\right ), 
\nonumber \\
\beta & \equiv & \ln\left( \frac{t_{\mathrm{w}}}{\tau_\infty} \right ).
\end{eqnarray}
Physically meaningful values of the parameters are in the
ranges 
$\alpha \approx 3$-4 (estimated from the measured $T_2$), $\beta \approx 8$-20 (corresponding to $\tau_\infty=10^{-8}$-$10^{-13}$~s).
  Eq.~(\ref{eq:amp}) may then be rewritten (in the limit $t_{\mathrm{w}} \gg \tau_{\mathrm{R}}$) as
\begin{eqnarray} 
\label{eq:ampR}
A(\tau)  =  \left (1-e^{-\frac{t_{\mathrm{w}}}{\tau}} \right )\Theta (\tau_{\mathrm{R}} -\tau) 
& + & e^{-\frac{t_{\mathrm{w}}}{\tau}} \nonumber \\
 =  \Theta (\tau_{\mathrm{R}} -\tau) & + & \Theta (\tau - \tau_{\mathrm{R}} ) e^{-\frac{t_{\mathrm{w}}}{\tau}} 
\nonumber \\
 \approx  \Theta (\tau_{\mathrm{R}} -\tau) & + & e^{-\frac{t_{\mathrm{w}}}{\tau}}.
\end{eqnarray}

We may now calculate the mean amplitude $ \langle A \rangle$ by averaging over the
energy $E$ with respect to the distribution of Eq.~(\ref{eq:distro}):
\begin{equation} %
\langle A \rangle = \int_0^\infty \mskip-20mu dE \mskip 8mu p(E) 
\left [\Theta (\tau_{\mathrm{R}} -\tau(E)) +  e^{-\frac{t_{\mathrm{w}}}{\tau}} \right ].
\end{equation} %
The first term in the integrand may be integrated exactly:
\begin{eqnarray} %
\langle A_1(T) \rangle & = & \int_0^\infty \mskip-20mu dE \mskip 8mu p(E)\, 
\Theta \left [ \tau_{\mathrm{R}} -\tau_\infty \exp\left (\frac{E}{T} \right ) \right ]\nonumber \\
& = & \int_0^{T\ln\frac{\tau_{\mathrm{R}}}{\tau_\infty}} \mskip-20mu dE \mskip 8mu p(E)\nonumber \\
& = & \frac{1}{2}\left [1 +\erf \left(\frac{\ln \frac{\alpha T}{E_{\mathrm{a}} }}{\sigma\sqrt{2}}  \right ) \right].
\end{eqnarray} %
For the second term, a closed expression can be obtained under the 
approximation $\exp\left[-\exp(-x)\right] \approx \Theta(x)$, yielding
\begin{eqnarray} %
\langle A_2(T) \rangle & = & \int_0^\infty \mskip-20mu dE \mskip 8mu p(E)\,
\exp\left (-\frac{t_{\mathrm{w}}}{\tau_\infty}e^{-\frac{E}{T}} \right )\nonumber \\
& \approx & \int_0^\infty \mskip-20mu dE \mskip 8mu p(E)\, 
\Theta (E-\beta T )\nonumber \\
& = & 1 - \int_0^{\beta T} \mskip-20mu dE \mskip 8mu p(E) \nonumber \\
& = & \frac{1}{2}\left [1 - \erf \left(\frac{\ln \frac{\beta T}{E_{\mathrm{a}}} }{\sigma\sqrt{2}}  \right ) \right].
\end{eqnarray} %
In summary, the result of the model is that the average amplitude may
be written
\begin{widetext}
 \begin{equation}
\label{eq:result}
\langle A(T) \rangle = 1 +  \frac{1}{2}\erf \left(\frac{\ln \frac{\alpha
      T}{E_{\mathrm{a}}} }{\sigma\sqrt{2}}  \right )  
-  \frac{1}{2}\erf \left(\frac{\ln \frac{\beta T}{E_{\mathrm{a}} }}{\sigma\sqrt{2}}  \right ).
\end{equation}
\end{widetext}

\subsection{Results}
\label{sec:results} 
The results of fitting the model to our NMR data are presented
below. The high- and low-temperature wipeout phenomena are treated 
separately, and points relative to the other temperature regime are rejected 
from the fits. Fitted values are shown in the panels.

\subsubsection{High temperature}

\begin{figure}[h]
\includegraphics[width=\columnwidth]{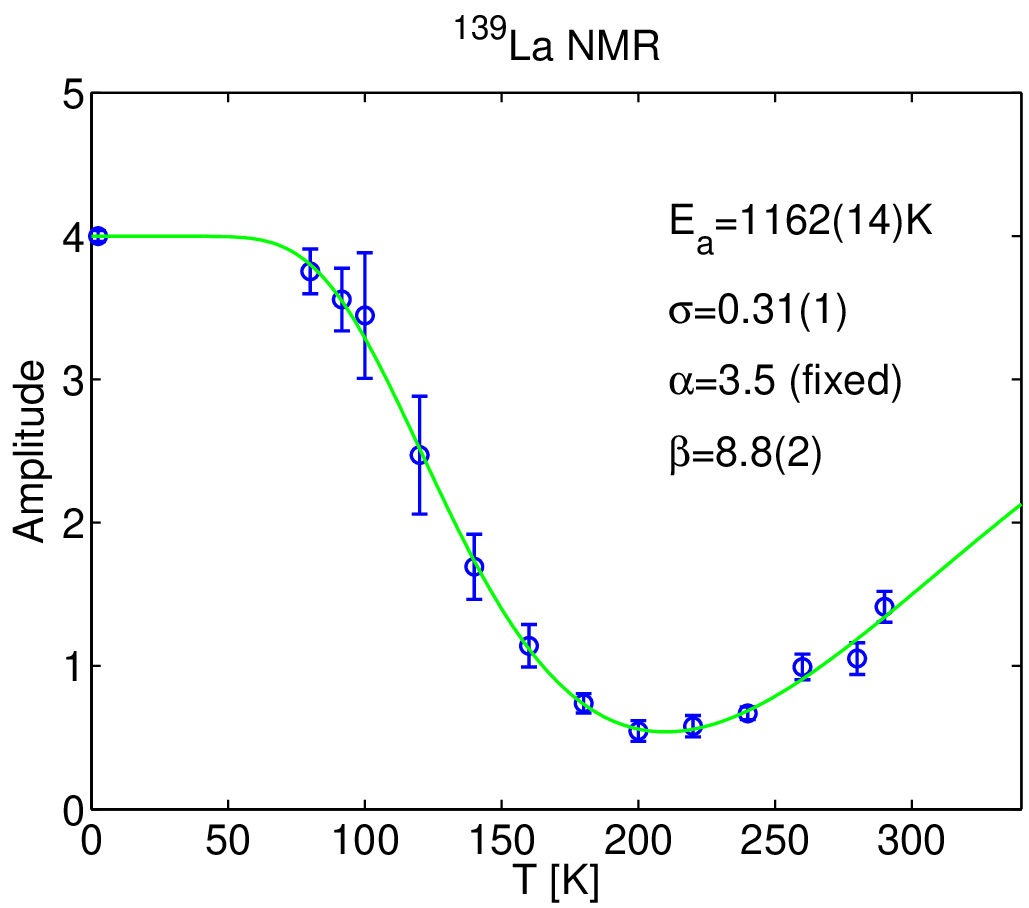}
\caption{\label{fig:LaHigh}
 {$^{139}$La NMR in the high temperature regime.
}}
\end{figure}

A satisfactory fit to Eq.~(\ref {eq:result}) can be obtained for the 
 $^{139}$La amplitude data [Fig.~\ref{fig:LaHigh}]. 
The parameter $\alpha$ governing the 
amplitude recovery at high temperature cannot be determined by the fit, due 
to the lack of data above room temperature, and hence it must be 
fixed. The estimate $\alpha=3.5$ corresponds to an asymptotic spin-spin relaxation 
rate $T_{2\,\infty}^{-1}\approx 4\times 10^{3}$~s$^{-1}$ at $T=\infty$, 
which is smaller than the room temperature value by a factor of 5 [see Fig.~4a of 
the main paper]. We note that the best-fit values of $E_{\mathrm{a}}$ and $\beta$
depend weakly on $\alpha$. 
For instance, fixing $\alpha=3$ and $\alpha=4$ yields $E_{\mathrm{a}}/k_{\mathrm{B}}=996(12)$~K, 
$\beta=7.5(2)$~K and $E_{\mathrm{a}}/k_{\mathrm{B}}=1328(16)$~K, $\beta=10.0(2)$~K, respectively,
and comparably accurate fits.
The orders of magnitude of $E_{\mathrm{a}}$ and $\beta$ are, however,
reasonably well established by these fits, in spite of the 
indeterminacy of $\alpha$. Notably, the above $\beta$ values correspond to 
infinite-temperature correlation times $\tau_{\infty}=10^{-9} - 10^{-8}$~s, 
in agreement with very slow hopping dynamics.

\begin{figure}[h]
\includegraphics[width=\columnwidth]{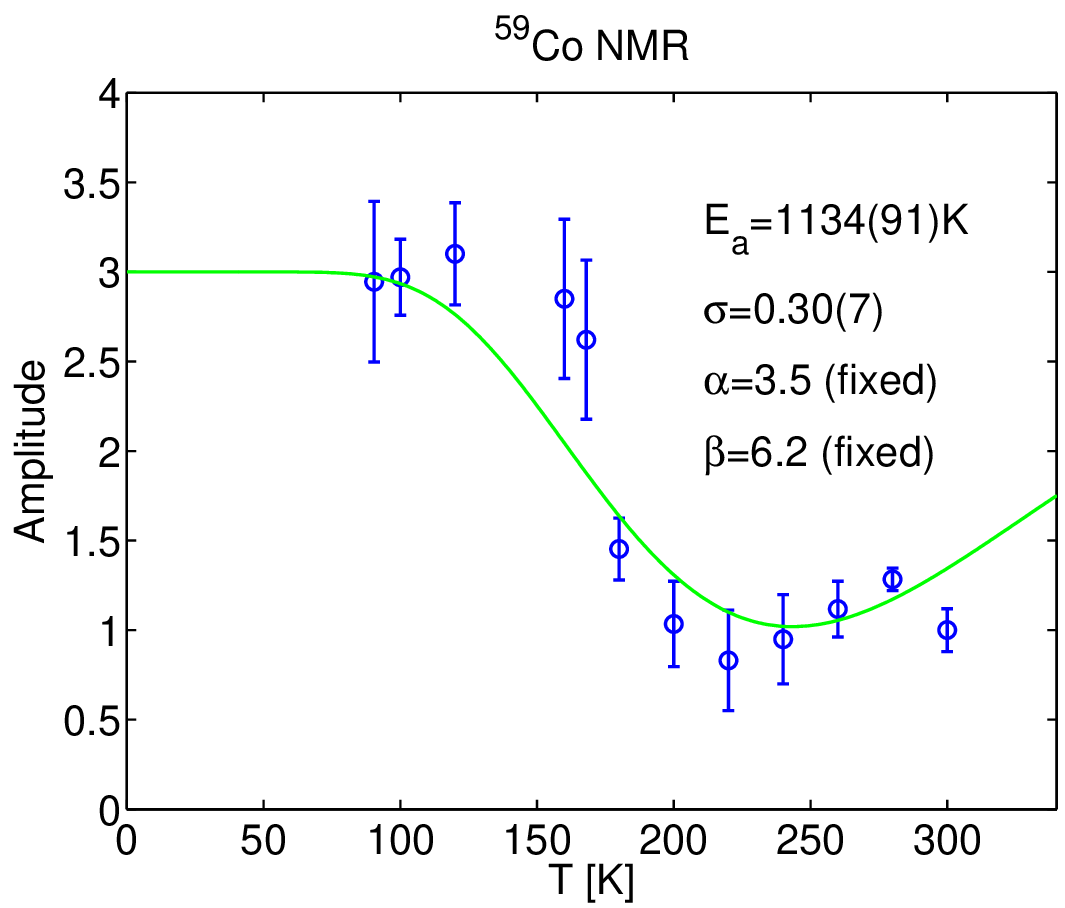}
\caption{\label{fig:CoHigh}
{$^{59}$Co NMR in the high temperature regime.
}}
\end{figure}

The temperature dependence of the $^{59}$Co signal amplitude agrees with that 
of 
$^{139}$La (Fig.~\ref{fig:CoHigh}). 
The poorer accuracy of the fit to Eq.~(\ref{eq:result}) seems to be due to 
the larger uncertainty on the 
spin echo amplitude, owing to the shorter $T_{2}$ of $^{59}$Co and the 
extrapolation of the spin echo at a delay $t=0$.
 
\subsubsection{Low temperature}
We apply our model only to $^{59}$Co here, which is the simplest probe of 
magnetism, since all Co$^{3+}$ sites are equivalent in the ideal case of a 
perfect  stripe order. 
[The analysis of $^{139}$La NMR amplitudes is complicated by the 
presence of 
two magnetically 
inequivalent La sites, one of which with a much smaller coupling to the 
Co$^{2+}$ moments.]

The application of Eq.~(\ref{eq:result}) to the $^{59}$Co amplitude vs.\ $T$ 
is further complicated by the overlap of the wipeout regime with the
magnetic transition at $T_{\mathrm{N}}\approx 35$~K. The nature of the spin 
fluctuations leading to the signal wipeout are most probably different in the 
paramagnetic and in the ordered phase, and the two cases must
therefore be treated 
separately. In both cases, either the $\alpha$ or $\beta$ parameters of the 
model, governing the 
low- or high-temperature stretch of the $A(T)$ curve, 
are irrelevant, as they are related to physically inaccessible regimes:
a virtual magnetically ordered phase at $T \gg T_N$ or a paramagnetic phase 
well below the $T_{2}^{-1}$ peak at $T\approx $ 20K.

A fit of the $^{59}$Co amplitude data at $T \le T_{\mathrm{N}}$,
with the additional constraint of a muon longitudinal relaxation peak
at a temperature  $T_{\mathrm{p}} \approx 18$~K [shown in Fig~2(b) of the
main article],
 is shown in Fig.~\ref{fig:CoLow}. 
Here $T_{\mathrm{p}}$ is implicitly defined via $\tau(T_{\mathrm{p}})^{-1} = 
\omega_{\mu}\approx 2\times 10^8$~s$^{-1}$  (see Eq.~(\ref{eq:activated}) and the article text).
With such a constraint, which guarantees the agreement between 
the NMR and $\mu^{+}$SR data, the parameter $\beta$ is a function of
the (fixed) parameter $T_{\mathrm{p}}$ 
and of $E_{\mathrm{a}}$. The parameter $\alpha$ is fixed to $\alpha  = 1$. The fit (5 
experimental points with 3 free parameters) 
yields 
$E_{\mathrm{a}}/k_{\mathrm{B}}=70(1)$~K and $\tau_{\infty}=7(1) \times 10^{-11}$~s$^{-1}$.   
We stress that the fitted activation energy is comparable with the analogous
quantity estimated in the paramagnetic phase (region I).


\commentout{longitudinal

\begin{figure}[h]
\includegraphics[width=\columnwidth]{139La_lowT}
\caption{\label{fig:LaLow}
{$^{139}$La NMR in the low temperature regime.
  }}
\end{figure}

}

\begin{figure}[h]
\includegraphics[width=\columnwidth]{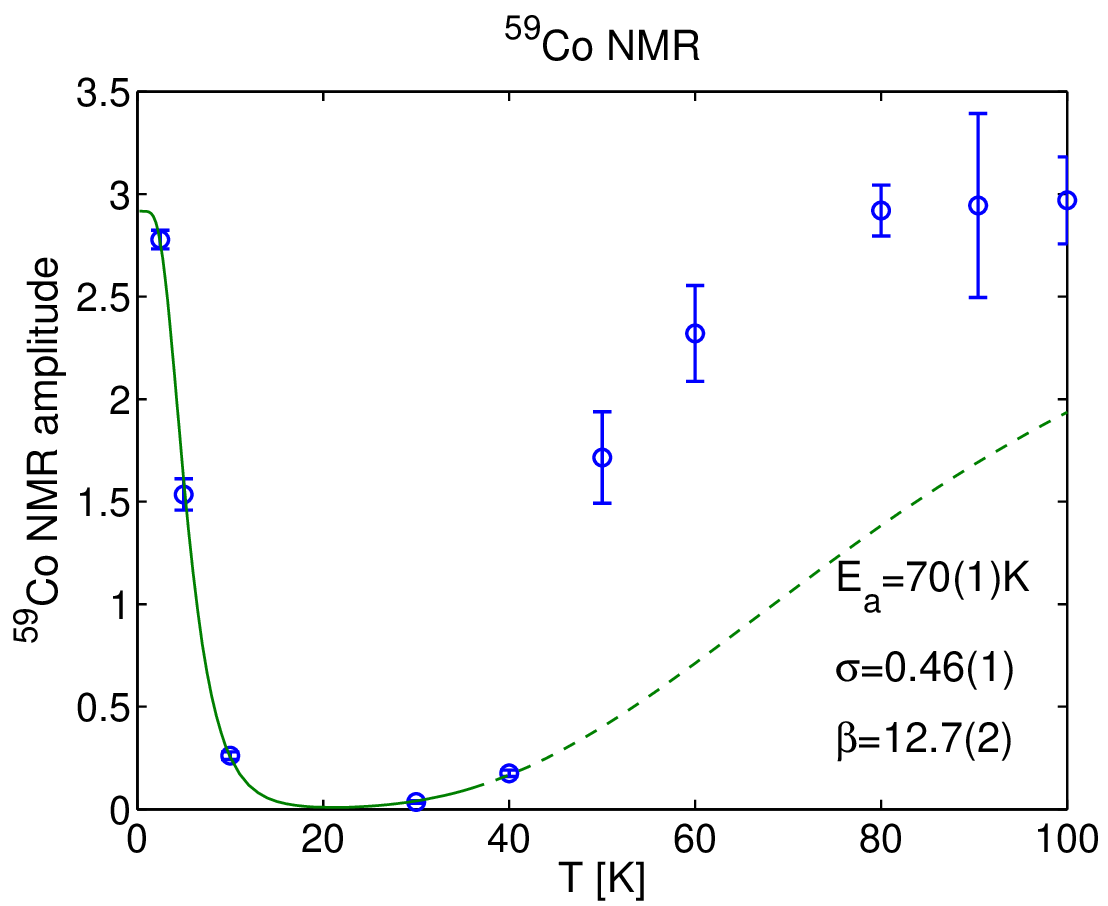}
\caption{\label{fig:CoLow}
{$^{59}$Co NMR in the low temperature regime.
}}
\end{figure}

\section{Muon spin relaxation}

In a muon-spin relaxation ($\mu^{+}$SR) measurement
spin-polarized positive muons are
stopped in a target sample. The positive muons are attracted to
areas of negative charge density and often stop at interstitial
positions. The observed property of the experiment is the
time evolution of the muon-spin polarization, the behavior
of which depends on the local magnetic field at the muon site.
Each muon decays with an average lifetime of 2.2~$\mu$s into
two neutrinos and a positron, the latter particle being emitted
preferentially along the instantaneous direction of the muon
spin. Recording the time dependence of the positron emission
directions therefore allows the determination of the spin
polarization of the ensemble of muons. In our experiments,
positrons are detected by detectors placed forward (F) and
backward (B) of the initial muon polarization direction.
Histograms $N_{\mathrm{F}}(t)$ and $N_{\mathrm{B}}(t)$ record the number of positrons
detected in the two detectors as a function of time following
the muon implantation. The quantity of interest is the decay
positron asymmetry function, defined as
\begin{equation}
a(t) = \frac{N_{\mathrm{F}}(t) -\alpha N_{\mathrm{B}}(t)}
{N_{\mathrm{F}}(t) + \alpha N_{\mathrm{B}}(t)},
\end{equation}
where $\alpha$ is an experimental calibration constant. The asymmetry
$a(t)$ is proportional to the spin polarization of the muon
ensemble.

\subsection{EMU data}

\begin{figure}
\begin{center}
\epsfig{file=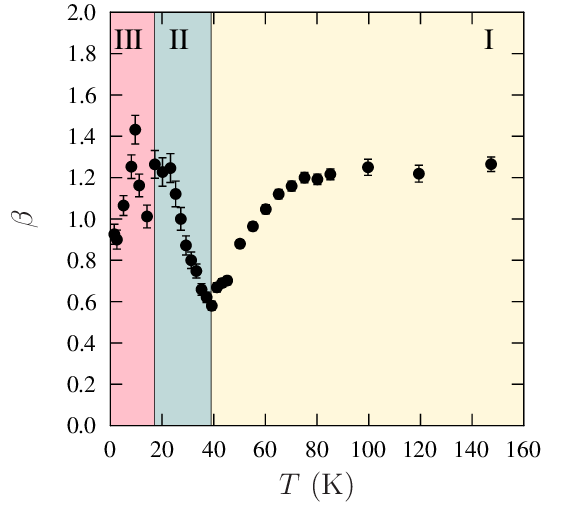,width=7cm}
\caption{The parameter $\beta$ from fits to data measured using the
  EMU instrument.\label{fig:emu_beta}}
\end{center}
\end{figure}

The evolution of the parameter $\beta$ from the fits, described in the
main text, to the data
measured using the EMU instrument at the ISIS facility are shown in
Fig.~\ref{fig:emu_beta}. This parameter also shows features at the
boundaries of the regions identified in the text with a minimum
observed on cooling to Region II and a maximum on entering region
III. It is worth noting that the ISIS data is dominated by the
limitations of the time window and so the stretched exponential
crudely models the response of two exponential relaxation components
expected on the grounds of the measurements made at S$\mu$S.



\subsection{DOLLY data}

\label{sec:muon}

The fitting of the DOLLY data involved a model which we describe here involving two
magnetically distinct muon sites. 
 The asymmetry $a(t)$ in zero field (ZF) $\mu^{+}$SR measurements may display both precessions
 and pure relaxations. 
Assuming that the muon spin at time $t=0$ lies along $\hat{z}$, the
axis of the detectors, an internal field $\boldsymbol{B}$ at the muon
site
causes a transverse precessing component,
$a_{\mathrm{T}}\exp(-\sigma^2 t^{2}/2) \cos \omega t $, 
(maximum amplitude $a_{\mathrm{T}}=a_{0}$ for $ \boldsymbol{B}\perp
\hat{z}$, $\omega=\gamma_{\mu} B$, $\gamma=851.4\,\mu$s$^{-1}$, 
field distribution width $\Delta =\sigma/(2\pi\gamma_{\mu})$) and a
longitudinal relaxing component, $a_{\mathrm{L}}\exp(-t/T_{1})$ 
(maximum amplitude $a_{\mathrm{L}}=a_{0}$ for
$\boldsymbol{B} \parallel \hat{z}$). 
A vanishing internal field, as occurs in the paramagnetic phase, 
results in $a_{\mathrm{T}}=0$ and consequently in the longitudinal
asymmetry 
$a_{\mathrm{L}}$ equalling the maximum value $a_{0}$. 

The model for the ZF muon asymmetry of \LSCO\ has two
transverse components
 ($a_{1\mathrm{T}}+a_{2\mathrm{T}}=a_{\mathrm{T}}$) and two
 longitudinal components
 ($a_{\mathrm{s}}+a_{\mathrm{f}}=a_{\mathrm{L}}$) reflecting to
 magnetically distinct muon sites in the material. 
The data is found to be well described with the following relaxation function
\begin{eqnarray}
a(t) &=& a_{\mathrm{s}} {\rm e}^{-t/T_{1\mathrm{s}}} + a_{\mathrm{f}}
{\rm e}^{-t/T_{1\mathrm{f}}}  + a_{1\mathrm{T}} {\rm
  e}^{-\sigma_{1}^{2} t^{2}/2} \cos \omega_1 t \nonumber \\
&+& a_{2\mathrm{T}} {\rm e}^{-\sigma_{2}^{2} t^{2}/2} \cos \omega_2 t.
\label{eq:asymmetry}
\end{eqnarray}
In an interval centered around the magnetic ordering transition the two
longitudinal terms are distinguished 
by a large ratio $T_{1\mathrm{s}}/T_{1\mathrm{f}}$. In the ordered phase the second
transverse term is over-damped 
($\sigma_{2}>\omega_{2}$) hence the cosine factor is set equal to
one. 

For the fitting routine, the temperature dependence of the eight free
parameters of Eq.~\ref{eq:asymmetry} was obtained firstly
in a fit of the complete data set. It suggested the following
constraints, leading to a minimal model of six free parameters below
$T_{\mathrm{N}}$ and three above: (i) a temperature independent
ratio $a_{\mathrm{s}}/a_{\mathrm{f}}$;  (ii) a total asymmetry
equal to the high temperature average $a_{0}$; (iii) a
temperature independent ratio $\sigma_{2}/\omega_{1}=0.75$. 
This model provides very good $\chi^{2}$ values, generally  within one 
standard deviation of that expected for the number of degrees of freedom, both on the entire time range  and on the early time range.